\documentclass[fleqn,10pt]{wlscirep}
\usepackage[utf8]{inputenc}
\usepackage[T1]{fontenc}
\usepackage{multicol}
\usepackage{amssymb}
\usepackage{latexsym}
\usepackage{gensymb}
\usepackage{textcomp}

\title{A hybrid machine learning/deep learning COVID-19 severity predictive model from CT images and clinical data}

\author[1,*]{Matteo Chieregato}
\author[2]{Fabio Frangiamore}
\author[3]{Mauro Morassi}
\author[4]{Claudia Baresi}
\author[1]{Stefania Nici}
\author[1]{Chiara Bassetti}
\author[2]{Claudio Bn\`a}
\author[1]{Marco Galelli}

\affil[1]{Fondazione Poliambulanza Istituto Ospedaliero, Unit of Medical Physics, Brescia, 25124, Italy}
\affil[2]{Tattile s.r.l., Mairano (BS), 25030 Italy}
\affil[3]{Fondazione Poliambulanza Istituto Ospedaliero, Department of Diagnostic Imaging, Unit of Radiology, Brescia, 25124, Italy}
\affil[4]{Fondazione Poliambulanza Istituto Ospedaliero, Information and Communications Technology, Unit of Lean Managing, Brescia, 25124, Italy}

\affil[*]{matteo.chieregato@poliambulanza.it}

\begin{abstract}
COVID-19 clinical presentation and prognosis are highly variable, ranging from asymptomatic and paucisymptomatic cases to acute respiratory distress syndrome and multi-organ involvement. We developed a hybrid machine learning/deep learning model to classify patients in two outcome categories, non-ICU and ICU (intensive care admission or death), using 558 patients admitted in a northern Italy hospital in February/May of 2020. A fully 3D patient-level CNN classifier on baseline CT images is used as feature 
 extractor. Features extracted, alongside with laboratory and clinical data, are fed for selection in a Boruta algorithm with SHAP game theoretical values. A classifier is built on the reduced feature space using CatBoost gradient boosting algorithm and reaching a probabilistic AUC of 0.949 on holdout test set. The model aims to provide clinical decision support to medical doctors, with the probability score of belonging to an outcome class and with case-based SHAP interpretation of features importance.
\end{abstract}
\begin{document}

\flushbottom
\maketitle
%
%
\thispagestyle{empty}
 
\section*{Introduction}
\label{sec1}
\begin{figure*}
\centering
\includegraphics[width=5 in]{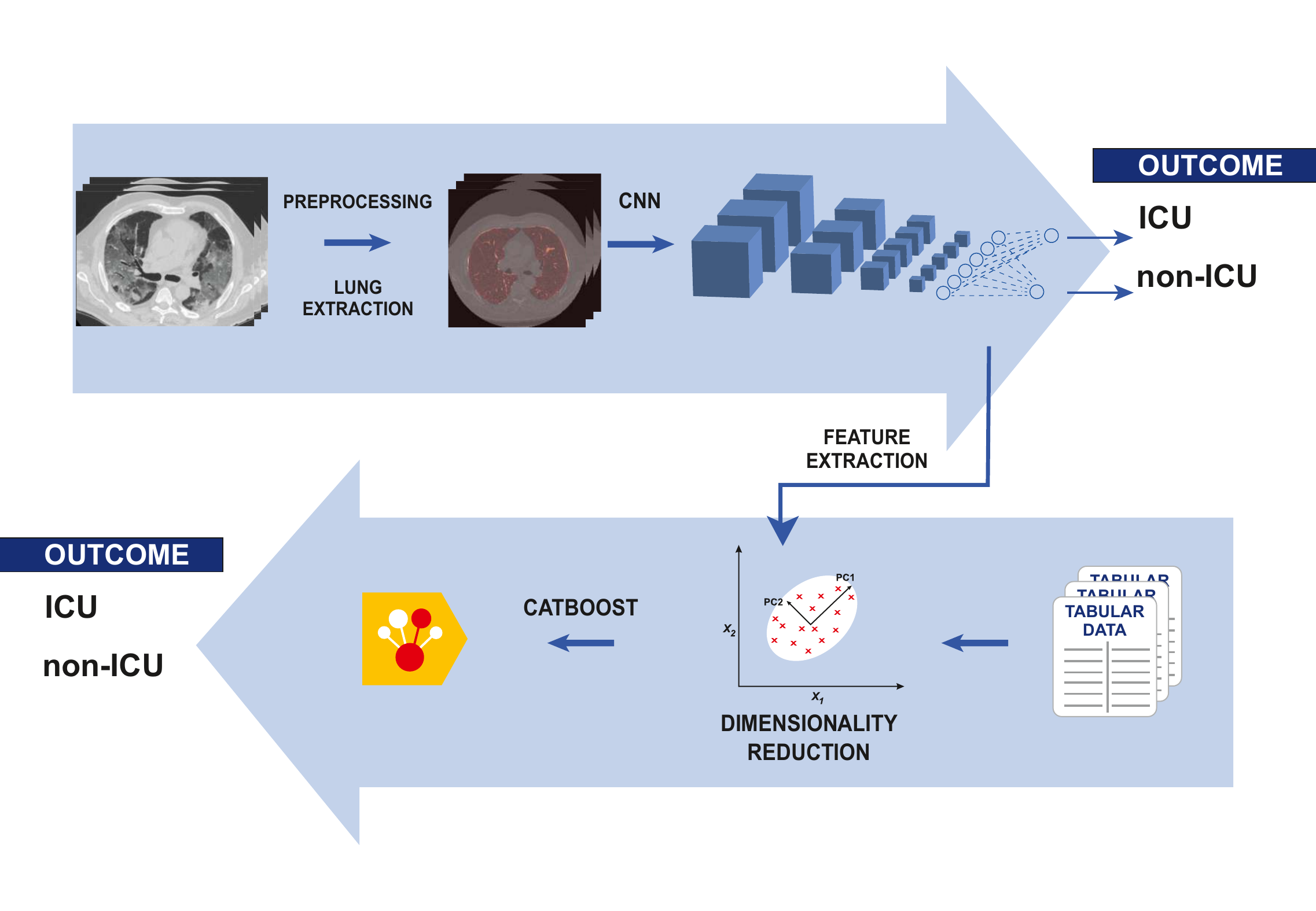}
\caption{A graphical representation of the proposed model.}
\label{fig:graphabs}
\end{figure*}

To date (May 2021),  more than one hundred millions of individuals have been reported as affected by COVID-19. More than two millions deaths have been ascribed to the infection. All over the world, the sheer numbers of the pandemic pose a heavy burden on emergency departments, hospitals, intensive care units and local medical assistance. From the beginning of the infection, it was apparent that COVID-19 encompasses a wide spectrum of both clinical presentations and consequent prognosis, with cases of sudden, unexpected evolution (and worsening) of the clinical and radiological picture \cite{struyf2020signs}. Such elements of variability and instability are still not fully explained, with an important role advocated for a multiplicity of pathophysiological processes  \cite{gupta2020extrapulmonary,li2020sars,wiersinga2020pathophysiology}. In this context, it would be natural to try to exploit techniques of artificial intelligence, fueled by the availability of large data amounts, to support clinicians. Indeed, a large number of efforts in this sense has
already been done, headed on different tasks, in particular diagnosis and
prognosis \cite{tayarani2020applications}. We focused on the latter, taking into account in particular
clinical usability.
We defined as our goal to build an hybrid machine learning/deep learning severity predictive model that can act as an auxiliary tool for patient risk assessing in clinical practice. 
In order to accomplish the objective, we considered essential the combination of
imaging and non-imaging data. We chose to exploit a Convolutional Neural Network (CNN) as feature extractor, and CatBoost, a last generation gradient boosting model, as classifier of tabular
data \cite{prokhorenkova2018catboost,dorogush2018catboost}.
The proposed model is represented graphically in Fig. \ref{fig:graphabs}.
The output of the model is both the percentage
score of the outcome and the SHAP (SHapley Additive exPlanations) 
evaluation of feature
importance in the individual
prediction \cite{lundberg2017unified, lundberg2018explainable}.
In this way, both synthetic and analytic information are provided to the
judgement of the clinician.
\section*{Methods}
\label{sec2}
\subsection*{Patients and dataset}
\label{subsec2.1}
\begin{figure}[htb]
\centering
\includegraphics[scale=.3]{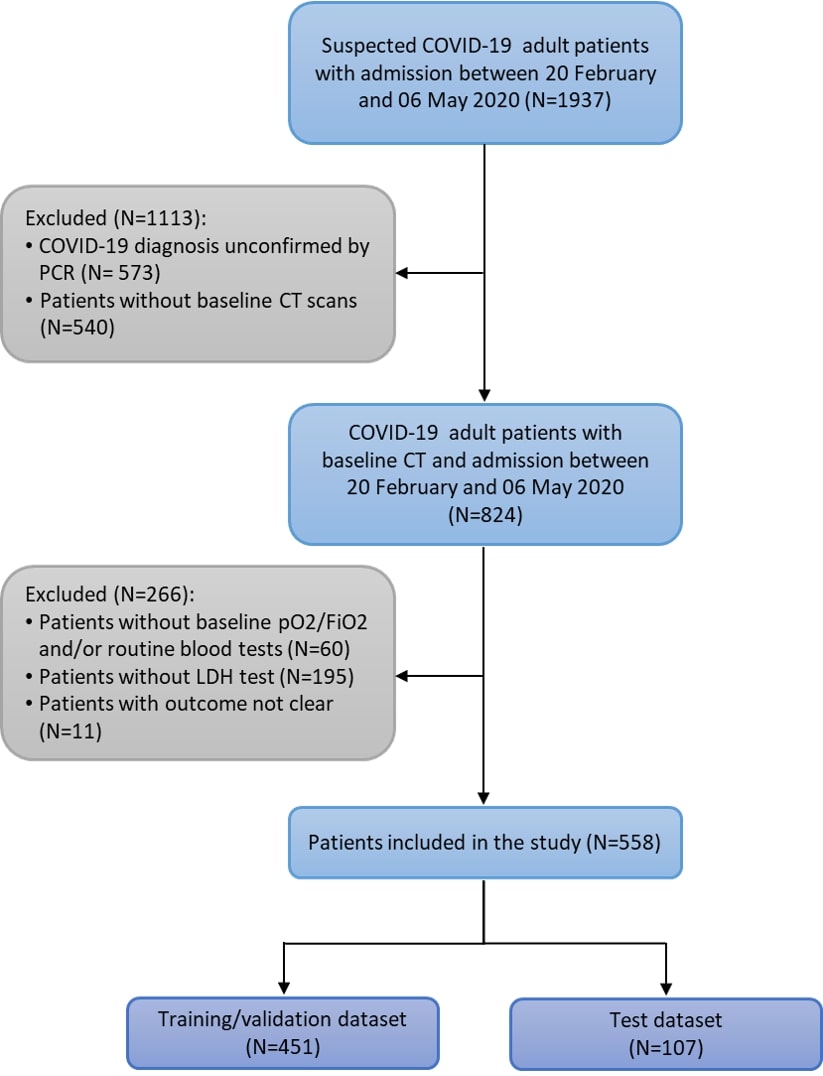}
\caption{Flowchart of patients inclusion/exclusion.}
\label{fig:flowchart}
\end{figure}

\begin{figure*}
\centering
\includegraphics[scale=0.3]{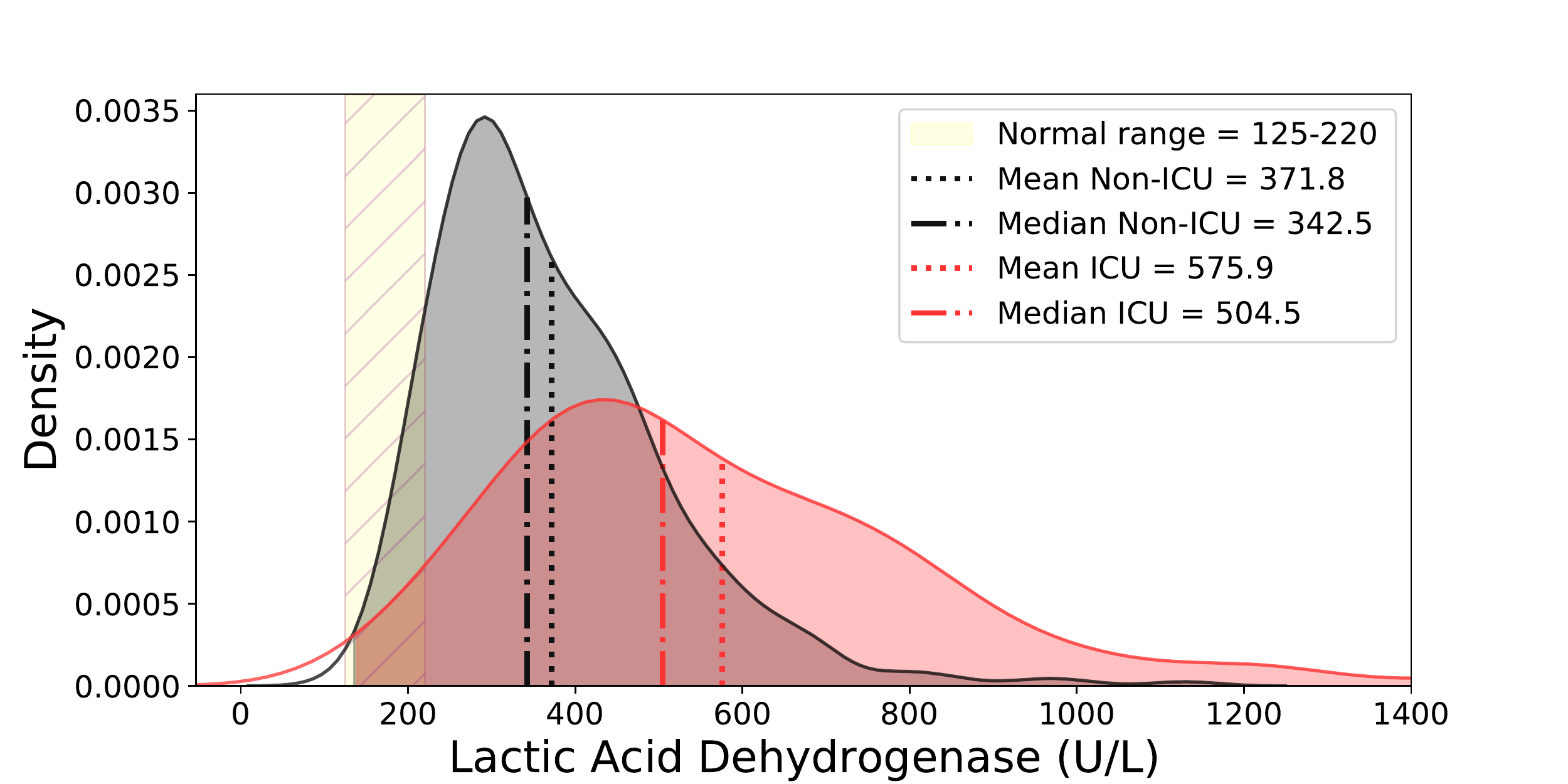}\quad\includegraphics[scale=0.3]{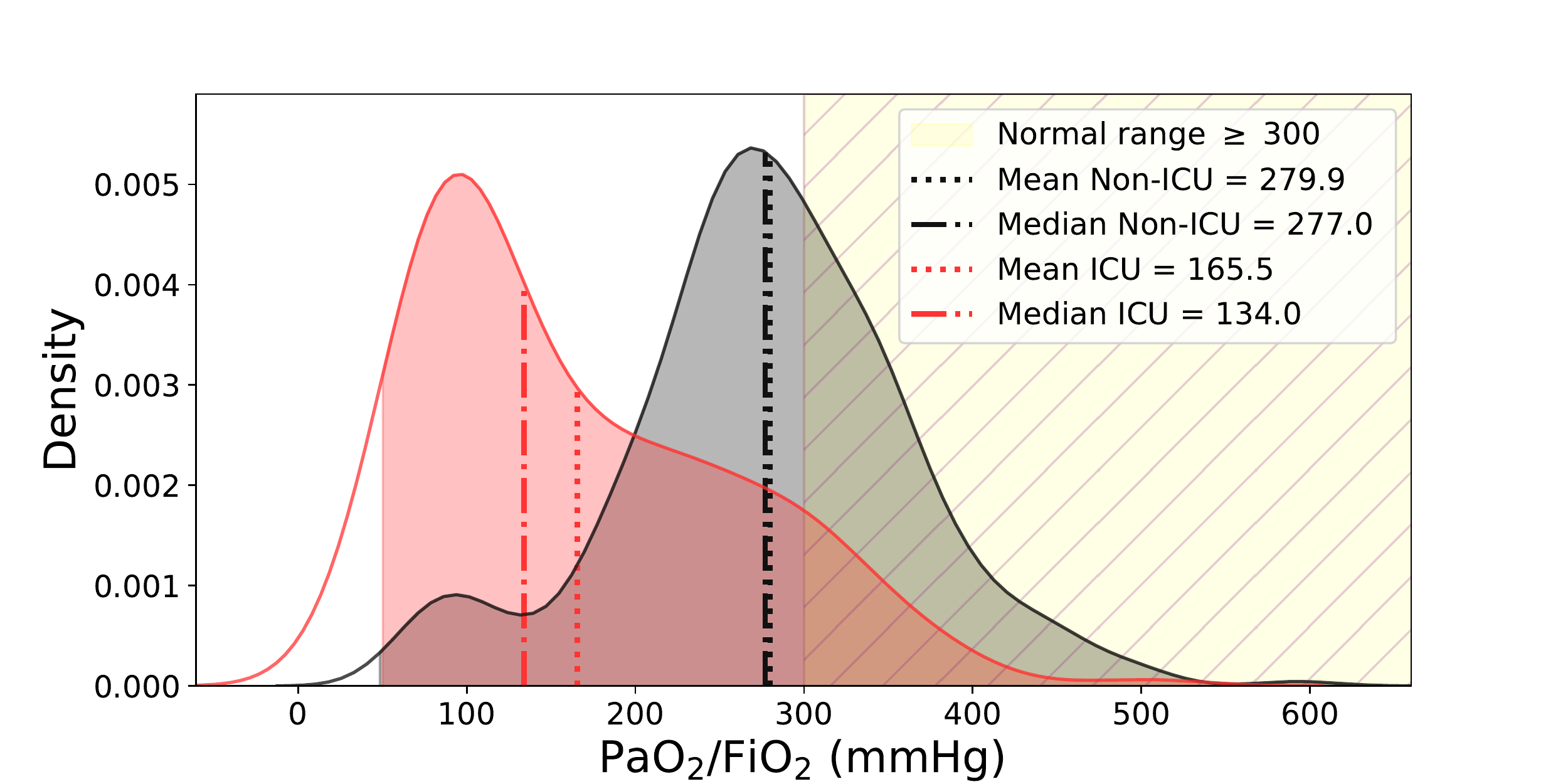}
\caption{Distributions of Lactic Acid Dehydrogenase and $ \rm{PaO_2/FiO_2}$ for patients in Non-ICU (gray) and ICU (red) severity classes. Yellow area is normal value range. Mean and median values are also indicated.}
\label{fig:distributionplots}
\end{figure*}

The dataset for this retrospective study consists of patients admitted to Fondazione Poliambulanza Istituto\\Ospedaliero (Brescia, Italy) between February 20, 2020 and May 6,
2020 with confirmed diagnosis of COVID-19. The hospital was at the forefront of fighting the disease outbreak in northern Italy in the first months of 2020. Diagnosis was made via nasopharyngeal swab analyzed through
 the Reverse Transcriptase-Polymerase Chain Reaction, RT-PCR.
Patients with baseline thoracic CT images,
 arterial blood gas analysis data, total blood counts and
 Lactate Dehydrogenase test (LDH) were considered for this study. This last has been  chosen as inclusion criterion due to his  effectivess as inflammatory biomarker for COVID-19 \cite{aloisio2020comprehensive,chen2020lactate}. 
We chose a binary outcome in two severity classes, evaluated at dimission, 
 defined as follows: 
\begin{enumerate}
    \item ICU class: death or intensive care unit admission;
    \item Non-ICU class: patients discharged as healed or transferred to non-COVID
      wards for further care.
\end{enumerate}
We excluded patients for which outcome reconstruction was uncertain (e.g. due to early transferral to other hospitals or care structures).
 A total of 558 patients met these criteria. Fig. \ref{fig:flowchart} shows the flowchart of patients selection. 
 Variables missing in more than 20\% of cases were excluded, even if their predictive efficacy has been advocated,
e.g. D-dimer \cite{lippi2020d}, Interleuchin-6 \cite{mcelvaney2020characterization}. Variables obviously redundant were merged (e.g. differential white cells count in
percent and absolute values). The 40
variables selected are shown in Table \ref{table:nonimagingdata}. Fig. \ref{fig:distributionplots} shows the respective
distribution of LDH and  $\rm {PaO_2/FiO_2}$  for both outcome classes. Deviations from normality are apparent for both classes.\\ The study has been approved by the ethical committee of Brescia (protocol number NP 4274 STUDIO GBDLCovid, session of 06/04/2021).
All methods were carried out in accordance with relevant guidelines and regulations. Informed consent collection or its waiving where not possible were conducted in agreement with the aforementioned protocol.

\begin{table*}

\caption{\label{table:nonimagingdata}Summary of clinical and laboratory variables used.}

\centering
\footnotesize
\resizebox{\textwidth}{!}{%

\begin{tabular}{|l|ccccc|}
\hline

\multicolumn{6}{|l|}{\bf{Binary}}\\
\hline

\hline

\multicolumn{2}{|l}{Variable}&   Total & Non-ICU & 
ICU &\\
\hline

\hline
Sex &&
F=166 M=392 &
F=121\ M=259 &
F=45\ M=133 &\\

Diabetes &&
Y=131\ N=427 &
Y=80\ N=300 &
Y=51\ N=127 &\\

Hypertension &&
Y=255\ N=303 &
Y=165\ N=215 &
Y=90\ N=88 &\\

Cardiovascular Disease &&
Y=263\ N=295 &
Y=164\ N=216 &
Y=99\ N=79 &\\

Oncological (last 5 yrs) &&
Y=41\ N=517 &
Y=33\ N=347 &
Y=8\ N=170 &\\

\hline
\multicolumn{6}{|l|}{}\\
\hline
\multicolumn{6}{|l|}{\bf{Numerical}}\\
\hline

\hline

Variable& Measure Unit&  Median & Median: Non-ICU & 
Median: ICU & Reference range\\
\hline

\hline
Age & yrs &
66 &
64 &
69 &\\
Body Mass Index (BMI) & & 
26 &
25.9 &
26 &\\

Body Temperature & \degree C&
37.5 &
37.4 & 
37.7 &
$<$ 37\\

Heart Rate (HR) & bpm &
92&
92 &
90 & 
60-100\\

Diastolic Blood Pressure (DBP) & mmHg &
76 &
77 &
75 & 
60-80\\

Systolic Blood Pressure (SBP) & mmHg &
127 &
127 &
127 & 
90-120\\

\hline
\multicolumn{6}{|l|}{\it{Arterial Blood Gas Analysis}}\\
\hline
$\rm pCO_2$  & mmHg &
36 &
35 & 
39 &
35-48\\

$\rm HCO_3$  & mmol/L &
25.4 &
25.4 &
25.4 &
21-28\\

$\rm PaO_2/FiO_2$  & mmHg &
255 &
277 &
134 &
$\geq$ 300\\

Lactate (LAC)  & mmol/L &
1.1 &
1.0 & 
1.3 & 
0.5-1.6\\

SO$_{2}$  & \% &
94 &
94.5 & 
91.8 & 
95-99\\

\hline
\multicolumn{6}{|l|}{\it{Complete Blood Count}}\\
\hline

White Blood Cell Count (WBC) & $\cdot 10^{9}/L$ &
7 &
6.6  & 
8.3 & 
4.5-10\\

Red Blood Cell Count (RBC) & $\cdot 10^{12}/L$ &
4.3 &
4.4 & 
4.3 & 
4.2-6.3\\

Hemoglobin (Hb) & g/dL &
13.1 & 
13.2 &
13 & 
14-18\\

Hematocrit (HCT) & \% &
39.8 & 
39.8 &
39.6 &
40-52\\

Red Blood Cell Distribution Width (RDW) & \% &
12.3 &
12.1 & 
12.6 & 
10.6-13.8\\

Granulocyte Neutrophils \% & \% &
78 & 
75 & 
84.7 & 
41-70\\

Granulocyte Eosinophils \% & \% &
0.2 & 
0.2 & 
0.2 & 
1-5\\

Granulocyte Basophils \% & \% &
0.2 & 
0.3 & 
0.2 & 
0.1-2\\

Monocytes \% & \% &
6.5 & 
7.4 & 
5 & 
1-12\\

Lymphocytes \% & \% &
14.2 & 
16.6 & 
9.4 & 
20-50\\

Platelets (PLT) & $ \cdot 10^{9}/L$ &
189 &
198 &
176 & 
130-450 \\

\hline
\multicolumn{6}{|l|}{\it{Additional Blood / Laboratory Analysis}}\\
\hline

Erythrocyte Sedimentation Rate (ESR) & mm/hour & 
5.5 &
5.4 &
5.6 & 
\it{variable}\\

C-reactive Protein (CRP)  & mg/L &
92 &
71 & 
151 & 
$<$ 5 \\

Albumin  & g/dL &
3.2 &
3.3 & 
3.2 & 
3.1-5.2\\

Prothrombin Time International Normalized Ratio (PT INR)  & & 
1 &
1 & 
1.1 & 
0.8-1.2\\

Aspartate Aminotransferase (AST)  & U/L &
46 & 
43 &
55 & 
$<$ 60\\

Alanine Aminotransferase (ALT)  & U/L &
34 &
33 & 
35 & 
$<$ 35\\

Total Bilirubin  & mg/dL & 
0.7 &
0.6 & 
0.7 & $<$ 1.2 \\

Creatine kinase (CK) & U/L &
102 &
86 & 
163 & 
30-200\\

Lactic Acid Dehydrogenase (LDH)  & U/L &
388 &
343 & 
505 &
125-220\\

Sodium & mmol/L &
140 &
140 & 
140 & 
136-145\\

Potassium  & mmol/L & 
4.1 &
4.1 & 
4.1 & 
3.3-5.1\\

Creatinine  & mg/dL & 
0.84 &
0.8 &
0.96 & 
0.72-1.18 \\

Urea  & mg/dL & 
38 & 
34 &
47 & 
18-55\\

\hline
\end{tabular}%
} 
    
\end{table*}

\subsection*{CT acquisition protocols}
\label{subsec2.2}
Chest CT were acquired using two 64 slices scanners Optima
CT 660 (GE Medical Systems, Milwaukee, USA). All patients were
examined in supine position. Four different acquisition protocols were used (see Table \ref{table:CTprot}). For all protocols, tube voltage was 120 kVp and automatic current modulation was used. The reconstruction algorithm were mixed filtered back projections-iterative (ASIR), usually with different proportions for the same acquisition (e.g. lung, bone and parenchyma optimized).

\begin{table}[ht!]
\caption{\label{table:CTprot} CT acquisition protocols}
\begin{tabular}{|l|cccc|}
\hline
Protocol & \#1&\#2&\#3&\#4 \\ 

\hline

\hline
 \% of cases  &  70 \%& 20 \%& 7 \%& 3 \% \\
 
 Transverse resolution (mm) & 0.765& 0.765& 0.765& 0.765\\
 
 Slice width (mm) &2.5&1.25&2.5&2.5\\
 
 Slice spacing (mm) &2.5&1.25&0.625&1.1\\
 
 Pitch &1.375&0.969&1.375&0.984\\
\hline
\end{tabular}
\end{table}

\subsection*{Image preprocessing}
\label{subsec2.3}
All CT scans were transformed with bicubic interpolation to a common spatial resolution of 1.625 mm $\times$ 1.625 mm $\times$ 2.5 mm. A rigid registration to a single CT picked as representative was performed in order to minimize small patient positioning differences. 
A lung mask was created on the basis of non-rigid method registration of a known CT with lung mask to the target CT \cite{sluimer2005toward,liauchuk2017imageclef}.
Once masked images were produced, a volume of size
160$\times$160$\times$240 was obtained with zero-padding. At this point,
different reconstructions for the same CT scans were merged (mean values were
used), obtaining
one single baseline volumetric image for each patient. The images were then z-normalized (subtracting means and dividing by standard deviation).

\subsection*{Tabular missing data}
\label{subsec2.4}
Non-imaging missing data have been replaced with median imputation. In order to avoid knowledge leakage, median imputation was always performed after test/ validation/ training split.

\subsection*{Model overview}
\label{subsec2.5}
The proposed model is composed as follows:
\begin{itemize}
    \item a fully 3D CNN patient-level classifier on CT images;
    \item feature extraction from the last Fully Connected Layer of the CNN;
    \item  a dimensionality reduction procedure including Principal Component Analysis (PCA) on extracted image features, a preliminary CatBoost model and the Boruta algorithm with the SHAP feature importance as metric (BorutaSHAP, \cite{ekeany_2020_4247618});
    \item  a CatBoost classifier on the reduced feature space.
\end{itemize}
The dataset was split in train/validation and test (holdout) subsets, in a 0.8:0.2 proportion (N$\rm{_{train/valid}}$=451, ICU=147, non-ICU=304 and N$\rm {_{test}}$=107, ICU=31, non-ICU=76, respectively; see Supplementary Table S1 for demographic data of the split). Ten fold stratified cross validation was applied in the train/validation set. The best model was then retrained on the joined training and validation sets and applied to the test
 set. Overall validation strategy is not trivial, due to feature extraction and feature selection steps. In brief, for each fold the validation set is first used for CNN hyperparameters choice and PCA analysis. Once image feature are extracted, we step back to the training set for the BorutaShap feature selection and a reduction of the CatBoost hyperparameter space to be searched. Validation set is again used to pick the best hyperparameter choice for the fold. 
 The output of the model is the percentage score of the classification and the SHAP feature importance values at patient level (see Fig. \ref{fig:CrossValidation}).\\
\subsection*{Volumetric convolutional neural network}
\label{subsec2.6}
\begin{figure*}
\centering
\includegraphics[width=\linewidth]{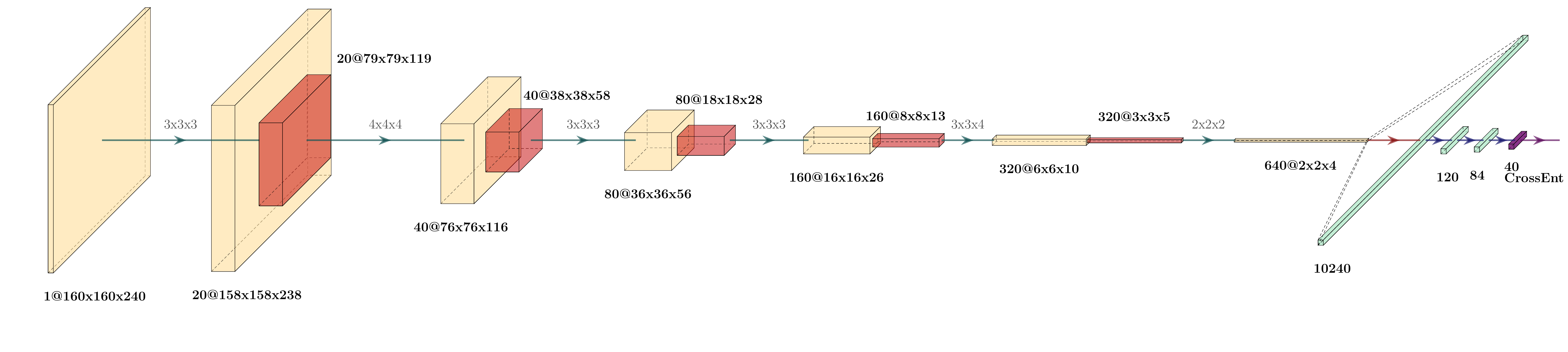}
\caption{A representation of the CNN architecture used. Actual model is volumetric, i.e. three spatial dimensions plus  a channels dimension. Green arrows represent convolution operations with stride of 1. A ReLU nonlinear activation is applied after convolutions, and then a 2x2x2 max pooling in order to reduce spatial dimensions. Red arrow represents flattening. Blue arrows are full connections (with a 0.25  dropout), purple arrow stands for the final classifier with Log SoftMax and Cross Entropy loss function.}
\label{fig:CNN}
\end{figure*}

The first block of the proposed model is a patient-level 3D CNN classifier, with six convolutional layers with ReLU activation followed by max pooling, and three fully connected layers with  a 0.25 dropout, plus a final classification layer. Group normalization is used, due to its better efficacy with small batches \cite{wu2018group}. The loss function is CrossEntropy.
The CNN block is shown in Figure \ref{fig:CNN}. 

\subsection*{CNN training and data augmentation}
\label{subsec2.7}
Data augmentation was performed in each fold on the fly, only for each training set, in the ten cross validation folds. 
Data augmentation techniques used were:
\begin{itemize}
    \item Affine deformation. During every epoch, there was a 50\% of probability to apply a random affine deformation with rotation between 0 to 10 degrees and a size variation up to 10\%.
    \item Elastic deformation. A random displacement was attributed to a grid of
      7$\times$7$\times$7 control points assigned to every images, with a
      maximum displacement equals to 10 voxels in each direction along
      cartesian axes. The displacement at every voxel was interpolated using a
      cubic B-spline function.
 \end{itemize}
All the techniques were implemented using the framework Torchio \cite{perez2020torchio}. Training was performed with the Stochastic Gradient Descent (SGD) optimizer and a fixed learning rate of $3\times 10^{-5}$. The number of epochs was chosen for each training/validation fold on the basis of AUC result on the validation set (the best in a fixed number of 50). For each fold, features at the input of the final classification layer were extracted (40 features). 

\subsection*{Dimensionality reduction}
\label{subsec2.8}
\begin{figure}[htb]
\centering
\includegraphics[width=4 in]{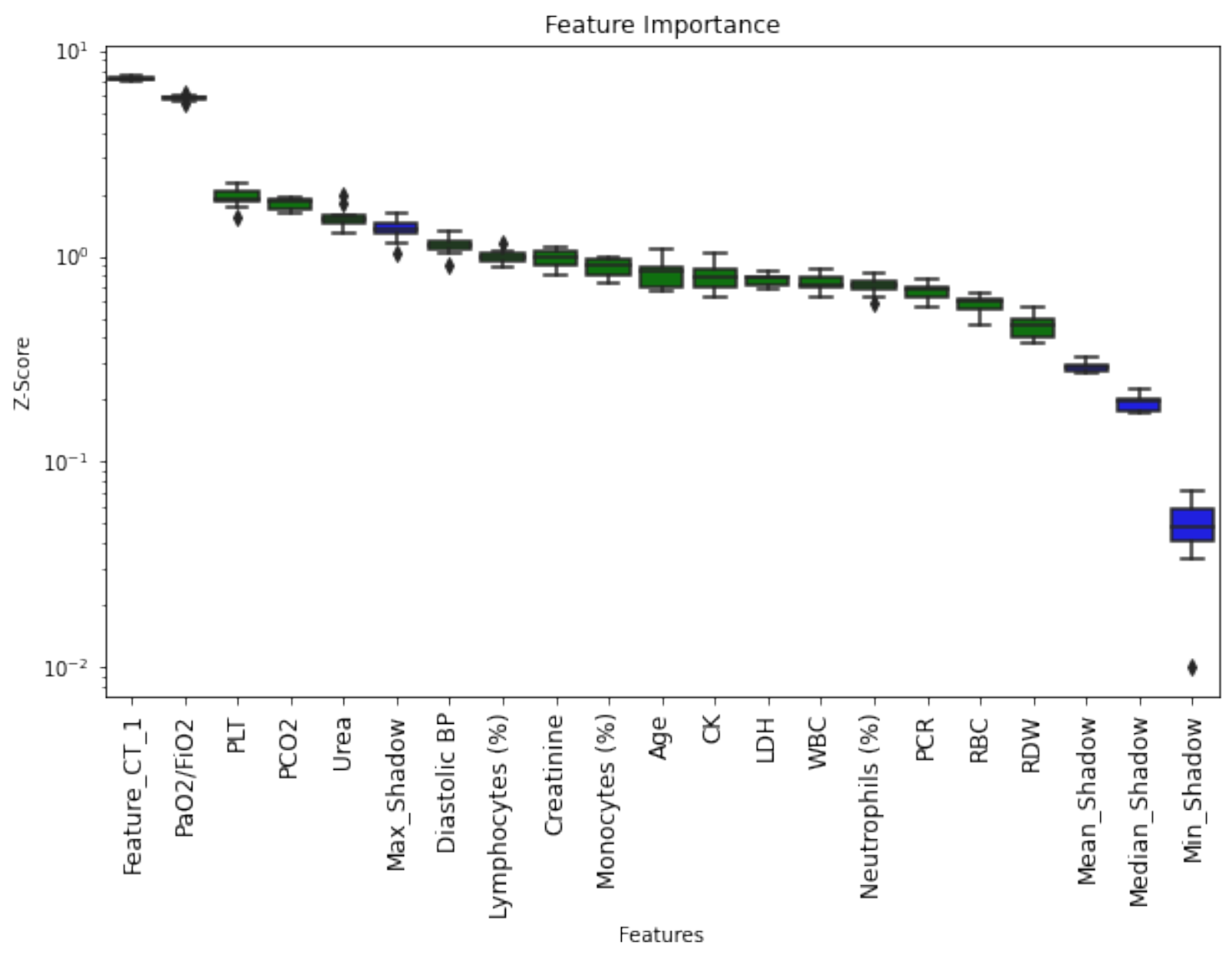}
\caption{A representative BorutaSHAP importance plot. Green are features to keep in the model for this fold. Blue are maximum, mean, median and minimum shadow features.}
\label{fig:borutashap}
\end{figure}

 A dimensionality reduction procedure was applied, articulated as follows:
\begin{itemize}
    \item Principal Component Analysis (PCA) of image extracted features (reduction of 40 extracted features to 5 principal components). 
    \item Training and hyperparameter optimization of a preliminary CatBoost classifier, with 40 non-imaging features and the 5 imaging features from PCA.
    \item A majority voting multi BorutaSHAP feature selection procedure.
\end{itemize}
All these steps were learned by the training subset, and then applied to validation (and test) subsets.
The usage of PCA to provide an out of the box, unsupervised, dimensionality reduction for CNN extracted features has been already proven effective in hybrid approaches \cite{lin2018convolutional}.  In this work we applied PCA only to CNN extracted image features, that can be considered natively agnostic, while the subsequent feature selection preserves interpretability.
The Boruta algorithm is an all relevant feature selection method, i.e. it tries to select all the features relevant for a given ensemble model. Relevance is evaluated against shadow features, that is dummy features created from real ones with random reordering of values \cite{kursa2010boruta}. 
 In the BorutaSHAP Python implementation, features and shadow features are compared by means of their SHAP importance values, producing therefore a result more consistent than other metrics \cite{lundberg2018consistent,hooker2019please}. 
The level of feature elimination can be tuned via a (percentile based) multiplicative factor on maximum shadow feature.
The ensemble model used was a preliminary CatBoost classifier, in which we fixed number of trees at 700 and learning rate at double of automatic CatBoost suggestion (so to to reach a quicker convergence).
For other hyperparameters, Bayesian optimization was performed with the automatic optimization framework Optuna 2.3.0 \cite{akiba2019optuna}, with 300 trials (0.8:0.2 calculation/evaluation split).\\
In our dataset, SHAP feature importance tends to have a slowly degrading distribution, except for the two most important features (CT first principal component and PaO$_{2}$/FiO$_{2}$; see Figure \ref{fig:borutashap}). Such a feature importance distribution can show some dependence on random picking of data sample.
 Therefore, in order to increase the robustness of feature selection,  we implemented a nested majority voting feature elimination strategy.\\
 The training subsample was divided in a 7:1 inner training and inner validation sets (with shuffling). A weak BorutaSHAP (85 percentile) feature selector with the preliminary CatBoost model was applied in each fold. Each time, feature importance was evaluated in the inner validation subset. Eight feature choices resulted.
A feature absent in six over eight choices was eliminated.
This procedure was applied to each of the ten training/validation folds (Figure \ref{fig:CrossValidation}). 

\subsection*{CatBoost model}
\label{subsec2.9}
\begin{figure}[htb]
\centering
\includegraphics[width=4.3 in]{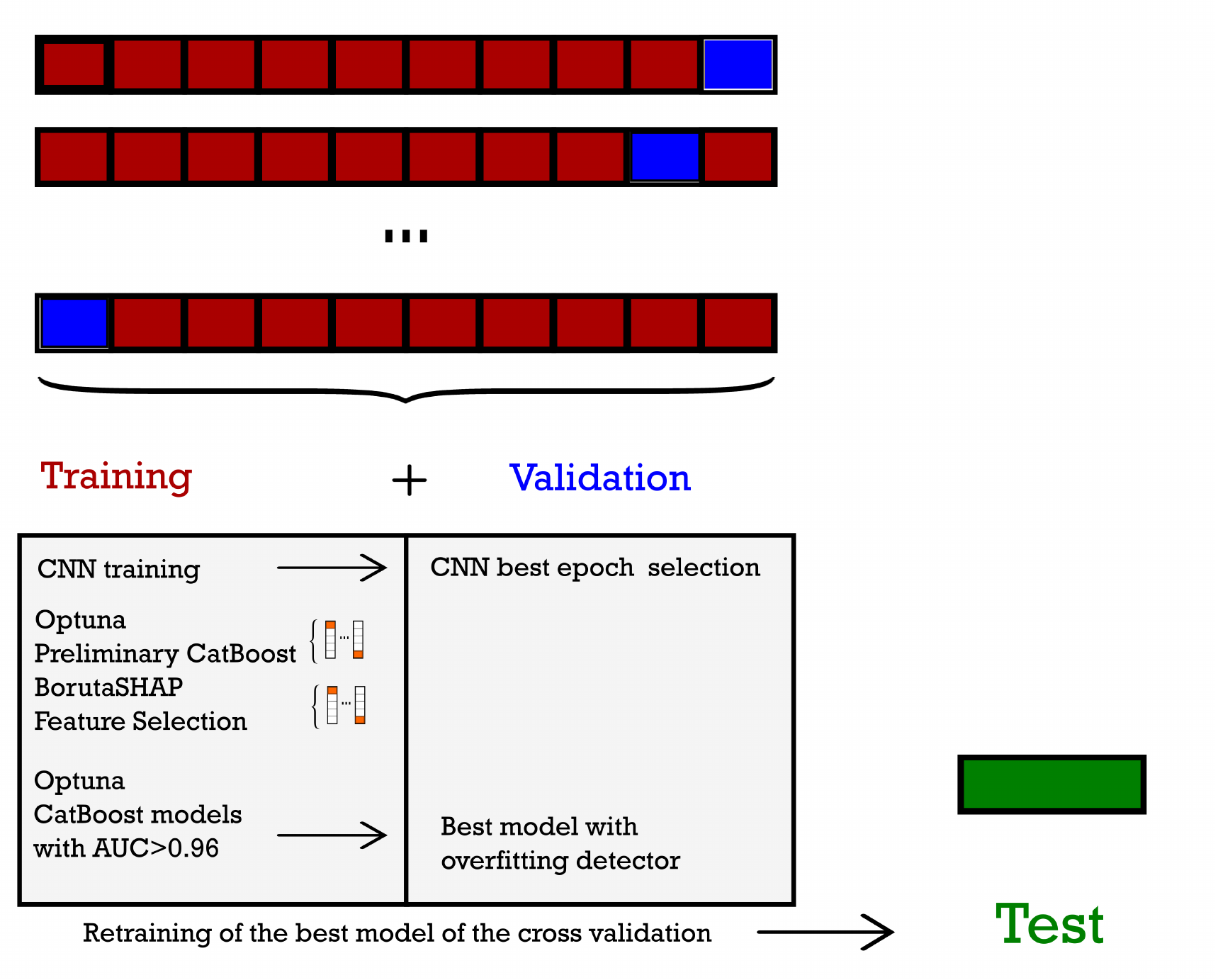}
\caption{A sketch of the cross validation procedure with dimensionality reduction. Both Optuna hyperparameter optimization and BorutaSHAP feature selection were computed with inner cross validation in the training set, not represented here.}
\label{fig:CrossValidation}
\end{figure}

We built a CatBoost classifier on the reduced feature set, using a two steps procedure for hyperparameters optimization:
\begin{itemize}
    \item Selection of a reduced number of hyperparameter combinations (the
best performing on the training set), with the aid of Bayesian optimization, at fixed learning rate and number of trees.
    \item The selected combinations were compared on the validation set, with a fixed learning rate and a number of trees optimized by the overfitting detector.
    \end{itemize}
The best model was chosen by AUC on its validation subset. It was then retrained on the joined training and validation subset, with a 120\% number of trees in order to keep in account the larger training size. Such final model was evaluated on the test/holdout dataset. A graphical resume of the cross validation and testing procedure is shown in Fig. \ref{fig:CrossValidation}.
The rationale of the procedure is to control the computational burden of hyperparameter search, and at the same time to fully exploit the potential of the overfitting detector for number of trees selection by means of early stopping. In the first step, Bayesian optimization in the training set was performed with the Optuna optimizer, with parameters as in the previous subsection. For the first step, learning rate was fixed at the values automatically calculated by CatBoost on the basis of the number of instances and features.
Models with AUC $\geq$ 0.96 were selected for validation testing (an empirically chosen threshold value). 
 In the second step, learning rate was fixed at a constant value of 0.008 (at the lower end of the range of values for the first step). The number of trees was picked with the CatBoost overfitting detector as the best performing on the validation subset, starting with a very large value, 20000. In this way, almost complete freedom is left to the overfitting detector to stop at the best iteration. In practice the final model has fixed learning rate, a Bayesian-optimized combination of hyperparameters, and a number of trees found by the overfitting detector. 
 Hyperparameters of the final CatBoost model are reported in Supplementary Table S2.

\subsection*{Implementation and code availability}
\label{subsec2.10}
The overall model implementation has been made in \\Python 3.7 with open source libraries. In particular the framework PyTorch 1.7 \cite{paszke2019pytorch} has been used for the CNN block. The PC utilized for the training is equipped with a Intel\textsuperscript{\tiny\textregistered} Core\textsuperscript{\tiny TM} i7-8700 CPU (6 cores, 12 threads, 3.2 GHz) and a NVIDIA\textsuperscript{\tiny\textregistered} GeForce\textsuperscript{\tiny\textregistered} RTX 2080 Ti GPU (11 GB memory).
The code is available at\\
 https://github.com/matteochieregato/GradientboostingCovid19. 
\section*{Results}
\label{sec3}
\subsection*{CNN results}
\label{subsec3.1}
\begin{figure}[htb]
\centering
\includegraphics[width=3.5 in]{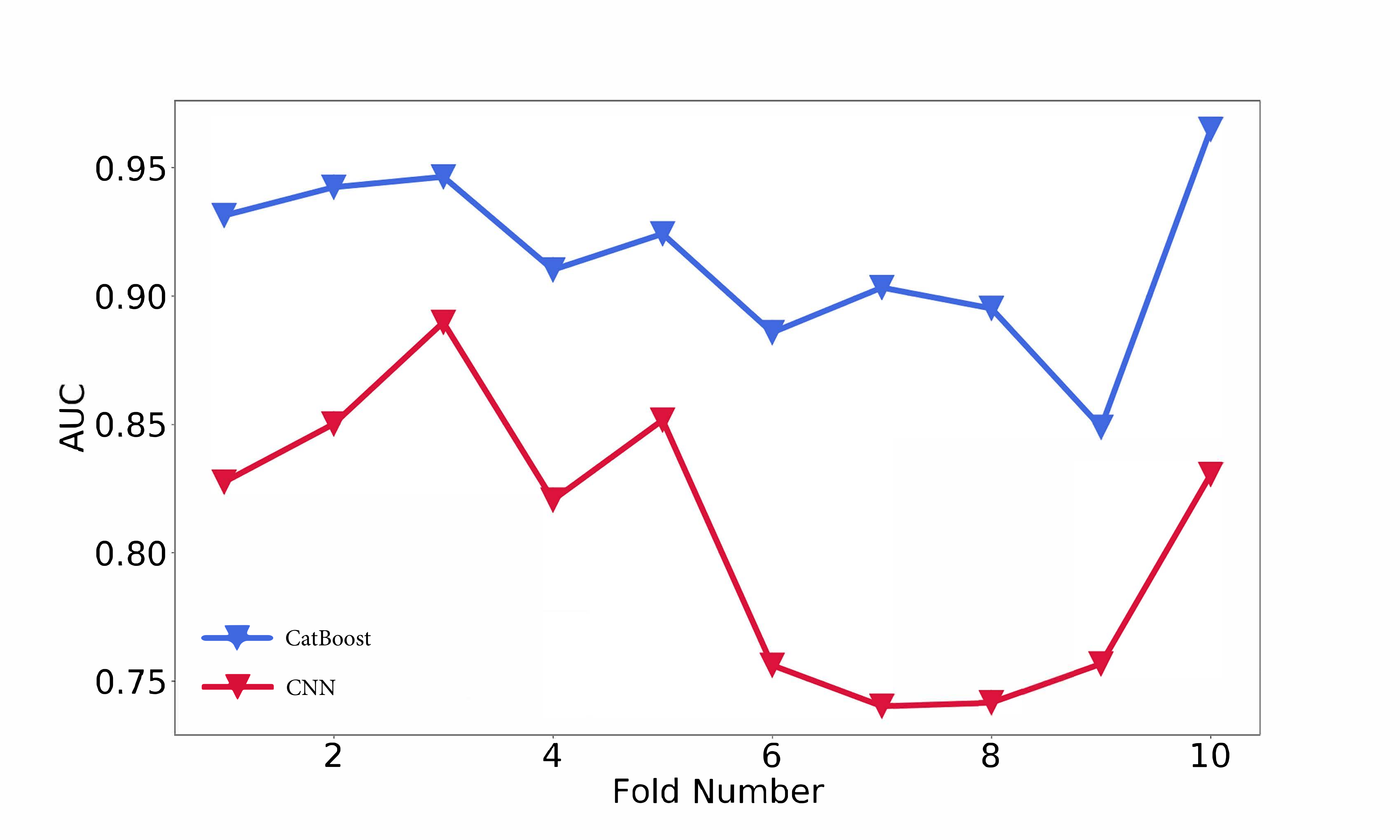}
\caption{Cross validation of the CatBoost and CNN classifiers. A roughly common trend can be discerned, however the highest score is reached at different folds (3rd for the CNN and 10th for the CatBoost classifier).}
\label{fig:CB_CrossVal}
\end{figure}

Results of the CNN classifier in terms of AUC is shown in Figure \ref{fig:CB_CrossVal} for the ten validation subsets. The third validation fold has the best AUC score, 0.889 (mean AUC in the ten folds is 0.806). Non-imaging feature selected by our procedure were consistent between folds. Variation in CNN results are likely to be imputed to the number of patients used for the training, not so large for deep learning. 
\subsection*{CatBoost predictive power}
\label{subsec3.2}
\begin{figure}[htb]
\centering
\includegraphics[width=3.5 in]{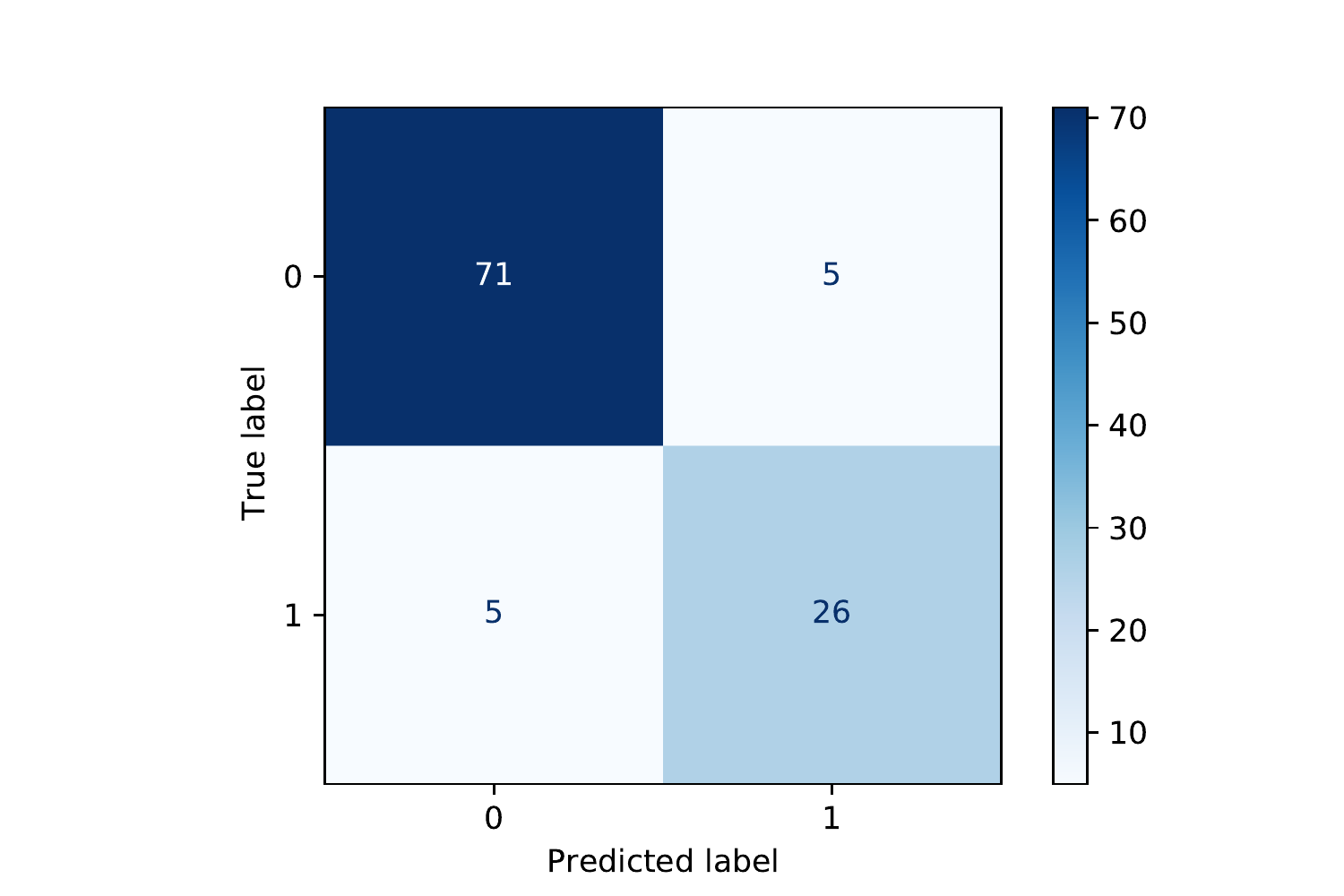}
\caption{Confusion matrix obtained with the best model on the test set (0 : non-ICU patients and 1 : ICU patients).}
\label{fig:ConfMatrix}
\end{figure}

Classification results of CatBoost in terms of AUC is shown in Fig. \ref{fig:CB_CrossVal} for each of the ten validation subsets (mean AUC = 0.915). The final best model reaches AUC = 0.949 in the test set, with a 95\% confidence interval of 0.899-0.986. The confidence  interval is calculated with the bootstrap method with 10000 folds resampling of the test set. Fig. \ref{fig:ConfMatrix} shows the confusion matrix for the test set (Sensitivity = 83.9\%, Specificity = 93.4\%). Since the model is intended as probabilistic classifier, it is optimized on probabilistic AUC, not on sensitivity and specificity. Setting the threshold for ICU prediction at 0.25 instead of 0.5, sensitivity becomes 90.3\% with a specificity of 89.5\%.

\subsection*{Feature selection and global level feature importance}
\label{subsec3.3}
\begin{figure}[htb]
\centering
 \includegraphics[width=3.7 in]{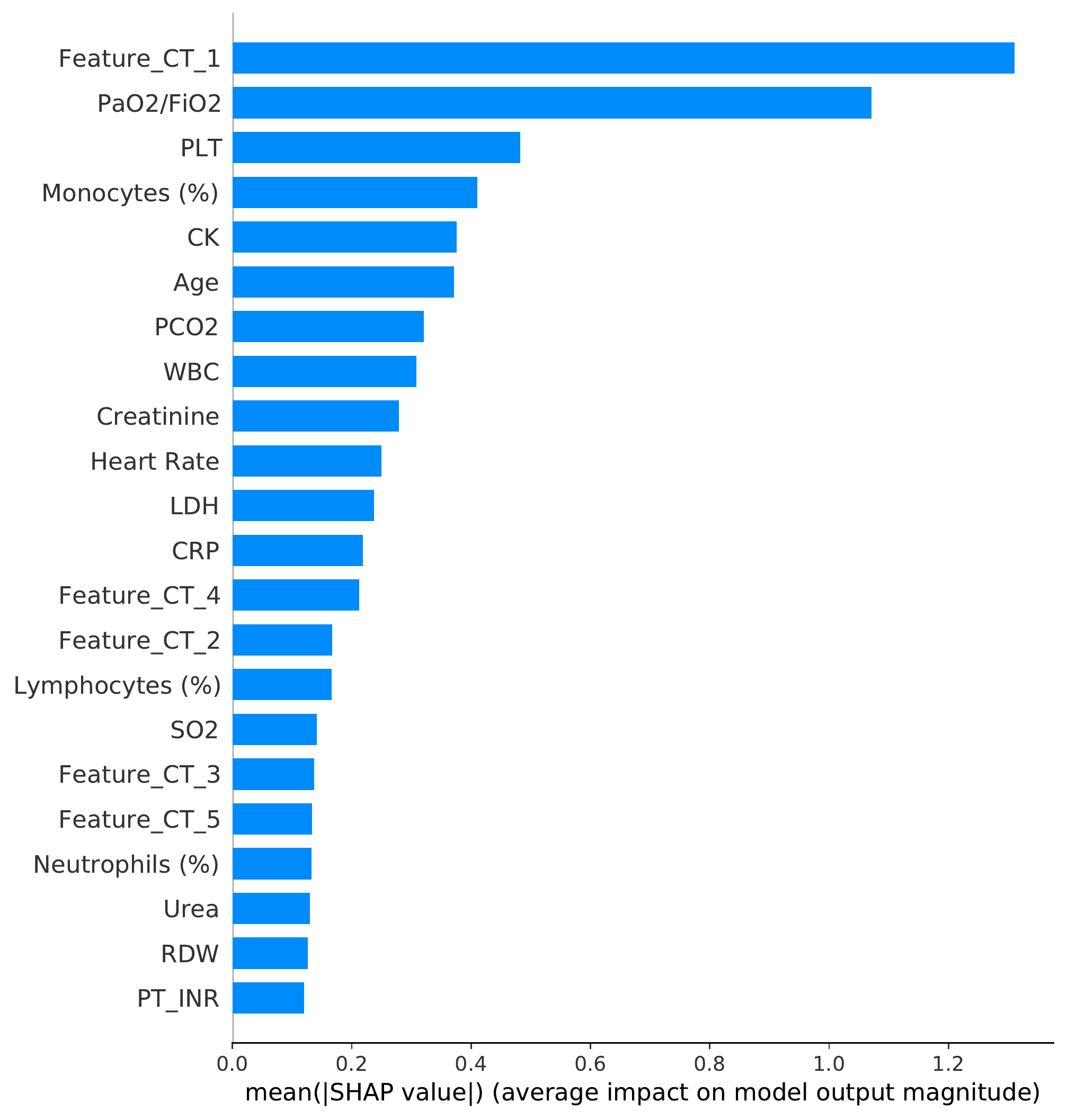}\par 
 \caption{Mean absolute value of the SHAP values for each feature in the test set.}
 \label{fig:GlobalImportance}
\end{figure}

Figure \ref{fig:GlobalImportance} shows the 22 features selected by our procedure in the best model, along with SHAP global feature importance in prediction over the test set.   
The first CT principal component and the $\rm PaO_{2}/FiO_{2}$ stand out.

\subsection*{Patient level feature importance}
\label{subsec3.4}
\begin{figure*}
\centering
\includegraphics[width=\textwidth]{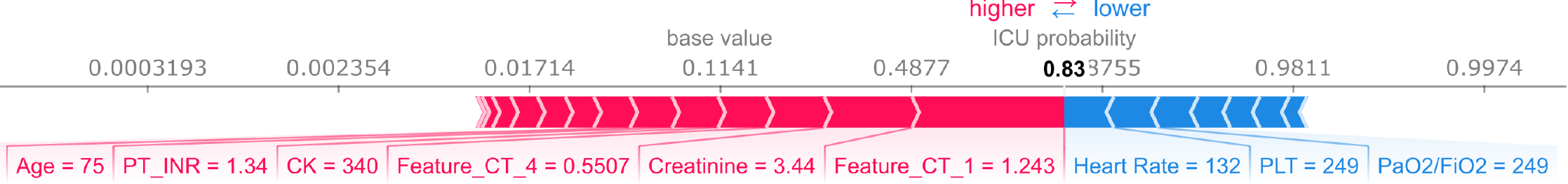}
\caption{Force plot of SHAP values for single patient. Less important features are omitted for the sake of visualization. 0.83 is the probability of ICU outcome. Features in red contribute positively to such probability, features in blue contribute negatively. More details are in Subsection Model introspection.}
\label{fig:SinglePatientImportance}
\end{figure*}

Figure \ref{fig:SinglePatientImportance} shows the SHAP output in terms of feature importance for a single prediction \cite{lundberg2018explainable}. Features are divided in colors and sides corresponding to the outcome to which they pull the prediction to (ICU is red and non-ICU is blue), bar size is which percentage score corresponds to the respective feature.
The case shown is correctly predicted as ICU with an 83\% score.
From the plot it is possible to see that  CT features (1 and 4) and most of parameters contribute to the ICU outcome prediction, despite a $\rm PaO_{2}/FiO_{2}$ ratio higher than most ICU cases (corresponding to mild Acute Respiratory Distress Syndrome, \cite{force2012acute}).

\subsection*{Model introspection}
\label{subsec3.5}
We analyzed on a case-by-case basis the patients for which the final model gave a wrong prediction, in particular ICU outcomes misclassified as non-ICU. It turned out that for 2 out of such 5 patients in the test set, there were meaningful additional information not taken in account by the model. In one case, there was a full scale D-dimer value (well known as indicator of poor outcome \cite{lippi2020d}). In the other, the patient is insulin-dependent type 1 diabetic. Diabetes comorbidity was eliminated by the feature selection procedure. Indeed, in our dataset a specific type 1 effect could have been hidden by the overwhelming majority of type 2 diabetic patients.
Such cases highlight the supporting role for which the proposed model is rightly intended for. Note that if these two cases were  to be excluded, sensitivity would be 89.7\%.
\section*{Discussion}
\label{sec4}
\subsection*{Imaging and non-imaging data combination}
\label{subsec4.1}
Complex tasks in clinic need
integration of radiological information with laboratory and clinical
information. Machine learning
methods are starting to be employed for such a purpose. 
Besides COVID-19 prognosis, examples can be Alzheimer disease classification and progress \cite{altaf2018multi} or the individuation of immunotherapy responders \cite{tunali2019novel}.\\ 
Radiological information is native as imaging data, while laboratory and
clinical information comes in
tabular form. 
Up to now, there is still no consensus on the best way to
combine these two types of data in machine learning models. 
In particular,
CNN are showing "unreasonable effectiveness" in image
related task \cite{LeCun,he2016deep} in the last years. However, the same is simply not
true for tabular data \cite{chui2018notes}, where ensemble models, and especially gradient
boosting variations (XGBoost, \cite{chen2016xgboost}, LightGBM, \cite{ke2017lightgbm}, CatBoost, \cite{prokhorenkova2018catboost}),  have the edge (\cite{kaggle2019}).\\ 
In principle, an integration of imaging and non-imaging information that harnesses the power of neural network in a combined model can be reached in a number of ways. Leaving apart simple CNN usage for segmentation \cite{zhang2020clinically}, they essentially boil down to three methods. First, combination of results of fully separated models \cite{ning2020open}. Second,
injection of non-imaging data, either as they are or after one or more fully connected layer, at some point of the CNN architecture, with a fully connected layer being the obvious choice \cite{meng2020deep, spasov2019parameter, gessert2020skin, liu2018joint}. Third, CNN as image feature extractor and a different machine learning model on the top to operate on both image extracted and non image features on equal footing (e.g. \cite{xu2020accurately}).\\
The first method is somewhat hampered by the fact that it can not keep in account any interaction between imaging and non imaging features.
The second method has the advantage of being end-to-end differentiable, and therefore trainable. As such, it is also more elegant to validate.
The third method can exploit a state-of-art model for heterogeneous data (e.g. gradient boosting, \cite{pang2019novel}, \cite{ren2017novel}, \cite{carvalho2020diagnosis} for extracted CNN features in XGBoost classifiers). The underlying machine learning architecture is less prone to data-starving, it is naturally integrable with advanced feature selection algorithms, and more readily explainable once agnostic features for images are accepted as such, since its symmetrical elaboration of non-imaging and image extracted features. \\
Our dataset consists of few hundred of patients,  a small number for CNN applications. Prognosis is a patient-level task, and as such we believe that number of patients, not of CT slices, is the fundamental number of instances. Furthermore, there is a perceived need for explainability of artificial intelligence applications, especially in the clinics (see below).
Therefore, we chose to sacrifice full differentiability and opted for the third method. As far as we know, this the first time that such a method to combine  CNN-extracted and non imaging data in a gradient boosting machine is used in a medical application. 

\subsection*{Model building and training}
\label{subsec4.2}
For COVID-19 prognosis, global features are likely to be more effective than spatially localized features
(that could be more useful for diagnosis in initial phases). Therefore a fully 3D
patient-level architecture is the more appropriate choice for the task. A CNN classifier allows to pick the high level representation features relevant to the task.
At the end of the network, a multiple fully connected layer structure allowed us a gradual reduction of the number of features before their extraction, so to balance it with non-imaging features.
CatBoost was used as the machine learning classifier for the final model. CatBoost is becoming increasingly applied in complex datasets \cite{hancock2020catboost}. It implements Ordered Boosting,
a permutation driven version of boosting algorithm, and Oblivious Decision
Trees, a particular type of decision trees (as well as other features we do
not treat here). Both should be especially effective in avoiding
overfitting. Hancock  and Khoshgoftaar \cite{hancock2020catboost} pointed out that CatBoost performance is likely
sensitive to hyperparameters choice. We especially picked by hand some
hyperparameters (Ordered Boosting as boosting type and Bayesian
bootstrap type) so to select the solution less prone to overfitting, using
Bayesian optimization for most of the others. The most influential
hyperparameters are the learning rate and the number of trees. For these,
CatBoost provides very powerful tuning methods, respectively with the
automatic learning rate estimate and the overfitting detector, and we made use
of both. 
The feature selection in our model is based on the combination of the Boruta algorithm with the SHAP metric, as implemented by Keany et al.  \cite{ekeany_2020_4247618}. 
The Boruta algorithm tries to find all relevant features for the task (and the model), not a compact subset that
minimize information loss for the classifier \cite{kursa2010boruta}. The use of the SHAP metric naturally keeps in account feature interactions and cooperative effects.
We implemented a majority voting procedure in order to exploit the strengths of BorutaSHAP, at the same time minimizing the risk of information loss and the dependence of subsampling randomness (Subsection CatBoost model). 
Since validation set is used as such both for CNN feature extractor and CatBoost hyperparameter choice, we can not completely exclude that some knowledge leaks from the feature extraction along our dimensionality reduction procedure up to the hyperparameter choice. 
We believe that our selected procedure, in particular the restriction of feature selection and Bayesian hyperparameter optimization on the training set should minimize the impact
of knowledge leakage (and therefore the risk of overfitting). In any case no leakage on the test set was possible, due to holdout from the start.

 \subsection*{Model interpretability}
\label{subsec4.3}
There is a general debate about the need
of interpretability of machine learning models for decision making \cite{bostrom2014ethics}. Notably, European Union legislation assesses the right to have an explanation of a decision made after automated data processing (GDPR16,
\cite{GDPR16}). We believe that an
even stronger push for model explainability comes from clinical needs. In
particular, an explainable model is not only more acceptable for doctors and patients, but becomes much more integrable with additional, out-of-the-model information (see Subsection Model introspection). In the proposed model, interpretability at global level and especially as single prediction level is given by the SHAP analysis. CT features, being extracted from the CNN classifier and the PCA reduction, are agnostic. However, one can still use them to appraise the overall weight of CT both in general and single case predictions. In particular, the first principal component is much more significant than the others, so it can be used as a proxy.

\subsection*{Limitations}
\label{subsec4.4}
There are limitations to the proposed model. First, the dataset comes from a
single center, in a localized period of time, with consequent trade-off
between data homogeneity and generalization power. 
Second, the number of
our patients is limited in comparison to the usual numbers in deep
learning classification tasks. Larger datasets naturally tend to reduce model
variance. To reduce the influence of these  two limitations, we took
particular care in trying to avoid overfitting.\\
Finally, any endpoint for COVID-19 related task can be potentially
influenced by the pressure posed to hospitals by the large numbers of patients
e.g. mortality rate and/or choice of admission to intensive care units can
change. We considered an ICU admission severity outcome to be   
more applicable in clinical context than a mortality prediction.
However, we are aware that such an outcome definition is
calibrated on our center (i.e. a different center can have different admission
criteria to intensive care unit). We believe that the choice of an interpretable,
probabilistic output can reduce the bias due to outcome choice.

\section*{Conclusion}
\label{sec5}

We built a COVID-19 prognostic hybrid machine-learning/deep learning model intended to be
usable as a tool that can support clinical decision making. 
The proposed model fully integrates imaging and non-imaging data. A 3D CNN classifier extracts patient level features from  baseline CT scans. A CatBoost classifier is applied on extracted features and laboratory and clinical data. Feature selection in the model is performed via the Boruta algorithm combined with the SHAP feature importance. Such architecture blends state-of-art machine learning for tabular data with the efficacy of a 3D CNN in building and selecting patient-level complex image features.
The tool is interpretable at global and at single
patient level, with the SHAP importance of features in obtaining the  percentage score of classification. Such analytical result is susceptible to be integrated by ulterior information that the clinician may have. We think that at the present state of things, this is the correct clinical usage of machine learning for COVID-19 prognostic tasks.
There are a handful of COVID-19 prognostic models that make use of radiological and clinical data with deep learning techniques, most of them either using deep learning only for segmentation \cite{zhang2020clinically,chao2020integrative, chassagnon2021ai} or keeping as a variable the resulting classification \cite{ning2020open}.
At odds with them, and more in line with \cite{meng2020deep} and \cite{xu2020accurately} (despite a much different architecture), the proposed method joins heterogeneous data at a lower representation level. As such, it allows models to take into account feature interactions. In particular an high degree of interaction between heterogeneous features is expected for COVID-19 prognosis task, due to complex relations between anatomical and functional lung involvement and systemic inflammatory response.\\
The proposed model was trained on a limited size dataset, without image segmentation from the radiologists. It would be therefore easily retrainable from scratch in order to adapt it to the mutable landscape of the pandemic, due to different variants of the virus, differences in the affected population demographics and effects of vaccine campaigns. Efforts in artificial intelligence triggered by the pandemic are likely to pave the way to future applications in different clinical contexts.  We believe that the integration of heterogeneous data and the interpretability of models will be keypoints for any clinical application involving complex tasks.

\section*{Data availability}
\label{sec6}
The dataset analyzed during the current study can be made available from the corresponding author on reasonable requests upon ethical comittee approval.

\bibliography{refs}

\section*{Acknowledgements}
The authors thank Maria Comi and Vittorio La Porta for their support to graphical abstract design and drawing, and Barbara Carli for useful insights about routine clinical practice.
\section*{Author contributions statement}

MC, MM and FF conceived the project. C Baresi and MM collected clinical data. SN and MC collected imaging data. MC and FF wrote the software and trained the model. MC, FF and SN analysed the results. MC, FF, SN, MM, C Bassetti and MG wrote the manuscript. MC, FF and SN prepared the figures. C Bn\`a and MG supervised the project. All authors revised the manuscript.

\section*{Competing interests}
The authors declare no competing interests.

\newpage
\setcounter{table}{0}
\renewcommand{\thetable}{S\arabic{table}}%
\setcounter{figure}{0}
\renewcommand{\thefigure}{S\arabic{figure}}%
\setcounter{section}{0}
    \renewcommand{\thesection}{S\arabic{section}}%

\section*{Supplementary Tables}

\begin{table*}[h!]
\caption{\label{table:nonimagingdatatv}Non imaging variables in the train/validation and test datasets}
\resizebox{\textwidth}{!}{%
\begin{tabular}{|l|cccc|}
\hline

\multicolumn{5}{|l|}{\bf{Binary}}\\
\hline

\hline

\multicolumn{2}{|l}{Variable}&   Train/validation & Test &\\
\hline

\hline
Sex &&
F=134 M=317 &
F=32\ M=75 &\\

Diabetes &&
Y=107\ N=344 &
Y=24\ N=83 &
\\

Hypertension &&
Y=199\ N=252 &
Y=56\ N=51 &
\\

Cardiovascular Disease &&
Y=222\ N=229 &
Y=41\ N=66 &
\\

Oncological (last 5 yrs) &&
Y=34\ N=417 &
Y=7\ N=100 &
\\

\hline
\multicolumn{5}{|l|}{}\\
\hline
\multicolumn{5}{|l|}{\bf{Numerical}}\\
\hline

\hline

Variable& Measure Unit& Median (10th-90th PCTL): Train/validation & 
Median (10th-90th PCTL): Test & Reference range\\
\hline

\hline
Age & yrs &
66 (50-80) &
64 (46.2-76.8) &
\\
Body Mass Index (BMI) & & 
25.7 (21.7-32.1) &
27.2 (23.4-33.3) &
\\

Body Temperature & \degree C&
37.5 (36.4-38.9)&
37.5 (36.3-38.8)& 
$<$ 37\\

Heart Rate (HR) & bpm &
92 (68-116)&
91 (74-118)&
60-100\\

Diastolic Blood Pressure (DBP) & mmHg &
76 (63-88)&
76 (64-89)&
60-80\\

Systolic Blood Pressure (SBP) & mmHg &
127 (107-158)&
129 (106-155) &
90-120\\

\hline
\multicolumn{5}{|l|}{\it{Arterial Blood Gas Analysis}}\\
\hline
$\rm pCO_2$  & mmHg &
36 (29-48)&
36 (31-43)& 
35-48\\

$\rm HCO_3$  & mmol/L &
25.4 (20.8-29.7) &
25.5 (22.2-29.1) &
21-28\\

$\rm PaO_2/FiO_2$  & mmHg &
254 (89-356) &
260 (86-378)&
$\geq$ 300\\

Lactate (LAC)  & mmol/L &
1.1 (0.6-2.2) &
1.0  (0.6-2.1)& 
0.5-1.6\\

SO$_{2}$  & \% &
94.1 (82.5-98.1) &
93.5 (80.7-98) & 
95-99\\

\hline
\multicolumn{5}{|l|}{\it{Complete Blood Count}}\\
\hline

White Blood Cell Count (WBC) & $\cdot 10^{9}/L$ &
7 (3.6-13.8) &
7.17   (3.8-12.8)&
4.5-10\\

Red Blood Cell Count (RBC) & $\cdot 10^{12}/L$ &
4.3 (3.6-5) &
4.4 (3.8-5) & 
4.2-6.3\\

Hemoglobin (Hb) & g/dL &
13 (10.8-14.8) & 
13.3 (11.2-14.7) &
14-18\\

Hematocrit (HCT) & \% &
39.7 (33.5-45.4) & 
39.9 (34.3-45.6) &
40-52\\

Red Blood Cell Distribution Width (RDW) & \% &
12.3 (11.5-14) &
12.0 (11.2-13.5) & 
10.6-13.8\\

Granulocyte Neutrophils \% & \% &
78 (57.4-90.3) & 
77.5 (54.6-89.3)& 
41-70\\

Granulocyte Eosinophils \% & \% &
0.2 (0.1-1.6)& 
0.2 (0.1-1.5) & 
1-5\\

Granulocyte Basophils \% & \% &
0.2 (0-0.9) & 
0.2 (0-0.9) & 
0.1-2\\

Monocytes \% & \% &
6.4 (2.8-11.5) & 
6.7 (3.5-11.9) & 
1-12\\

Lymphocytes \% & \% &
14.1 (5.4-30.4) & 
14.2 (5.8-31.8) & 
20-50\\

Platelets (PLT) & $ \cdot 10^{9}/L$ &
181 (107-335) &
204 (126-309) & 
130-450 \\

\hline
\multicolumn{5}{|l|}{\it{Additional Blood / Laboratory Analysis}}\\
\hline

Erythrocyte Sedimentation Rate (ESR) & mm/hour & 
5.5 (1.9-8.1)&
5.4 (1.9-8)&
\it{variable}\\

C-reactive Protein (CRP)  & mg/L &
94.3 (17-246.9) &
82.7 (19.1-229) & 
$<$ 5 \\

Albumin  & g/dL &
3.2 (2.7-3.7)&
3.2 (2.8-3.6) & 
3.1-5.2\\

Prothrombin Time International Normalized Ratio (PT INR)  & & 
1.02 (0.92-1.27) &
1 (0.93-1.19)& 
0.8-1.2\\

Aspartate Aminotransferase (AST)  & U/L &
46 (25-101) & 
48 (23-103) &
$<$ 60\\

Alanine Aminotransferase (ALT)  & U/L &
33 (13-92) &
38 (14-107) & 
$<$ 35\\

Total Bilirubin  & mg/dL & 
0.7 (0.5-1.3)&
0.6 (0.4-1.2)&
$<$ 1.2 \\

Creatine kinase (CK) & U/L &
102 (33-498)&
100 (36-363) & 
30-200\\

Lactic Acid Dehydrogenase (LDH)  & U/L &
391 (229-699) &
375 (242-699) & 
125-220\\

Sodium & mmol/L &
140 (135-144) &
140 (136-145)& 
136-145\\

Potassium  & mmol/L & 
4.1 (3.4-4.9) &
4 (3.5-4.9) & 
3.3-5.1\\

Creatinine  & mg/dL & 
0.85 (0.64-1.6) &
0.83 (0.68-1.6) & 
0.72-1.18 \\

Urea  & mg/dL & 
38 (21-82) & 
38 (24-90)&
18-55\\

\hline
\end{tabular}
}
\end{table*}
\pagebreak
\begin{table*}[h!]
\caption{\label{table:catboosthyper} CatBoost main hyperparameters in the final trained model}
\centering
\footnotesize
\begin{tabular}{|l|cc|}
\hline

\multicolumn{2}{|l|}{Hyperparameter}& Value\\
\hline

\hline

\hline
Loss function  &&
LogLoss\\
Iterations/number of trees  &&
12483\\
Learning rate  &&
0.008\\
L2 leaf regularization term &&
4.136\\
Boosting type&&
Ordered
\\
Bootstrap type&&
Bayesian\\
random subspace method &&
0.076\\
Bagging temperature
&&
3.102\\

\hline
\end{tabular}

\end{table*}

\end{document}